\documentclass[manuscript,screen]{acmart}

\AtBeginDocument{%
  \providecommand\BibTeX{{%
    \normalfont B\kern-0.5em{\scshape i\kern-0.25em b}\kern-0.8em\TeX}}}

\setcopyright{acmcopyright}
\copyrightyear{2023}
\acmYear{2023}
\acmDOI{XXXXXXX.XXXXXXX}

\acmConference[ICMLT 2023]{International Conference on Machine Learning Technologies (ICMLT 2023)}{March 10-12, 2023}{Stockholm, Sweden}

\begin{document}

\title{Sleep Quality Prediction from Wearables using Convolution Neural Networks and Ensemble Learning}

\author{Ozan Kılıç}
\authornote{Both authors contributed equally to this research.}
\email{ozaankilc@gmail.com}
\orcid{0000000207159466}
\author{Berrenur Saylam}
\authornotemark[1]
\email{berrenur.saylam@boun.edu.tr}
\orcid{0000000282948342}
\author{{\"O}zlem Durmaz İncel}
\email{ozlem.durmaz@boun.edu.tr}
\orcid{0000000262297343}
\affiliation{%
  \institution{Computer Engineering Department, Boğaziçi University}
  \city{İstanbul}
  \country{Turkey}
  \postcode{34342}
}

\renewcommand{\shortauthors}{Kılıç, et al.}

\begin{abstract}
 Sleep is among the most important factors affecting one's daily performance, well-being, and life quality. Nevertheless, it became possible to measure it in daily life in an unobtrusive manner with wearable devices. Rather than camera recordings and extraction of the state from the images, wrist-worn devices can measure directly via accelerometer, heart rate, and heart rate variability sensors. Some measured features can be as follows: time to bed, time out of bed, bedtime duration, minutes to fall asleep, and minutes after wake-up. There are several studies in the literature regarding sleep quality and stage prediction. However, they use only wearable data to predict or focus on the sleep stage. In this study, we use the NetHealth dataset, which is collected from 698 college students' via wearables, as well as surveys. Recently, there has been an advancement in deep learning algorithms, and they generally perform better than conventional machine learning techniques. Among them, Convolutional Neural Networks (CNN) have high performances. Thus, in this study, we apply different CNN architectures that have already performed well in the human activity recognition domain and compare their results. We also apply Random Forest (RF) since it performs best among the conventional methods. In future studies, we will compare them with other deep learning algorithms.
\end{abstract}



\received{30 January 2023}
\received[accepted]{10 February 2023}

\maketitle
\section{Introduction}
Understanding the underlying factors of sleep quality may significantly improve people's living standards. There could be many factors that need to be discovered that affect sleep quality metrics. These factors can be exposed with the help of big data and accurate deep-learning models. This study aims to find meaningful relations between sleep measurements with comprehensive multi-modal data, including personal surveys and everyday wearable device data.

All the mentioned data are derived from the NetHealth open-source dataset \footnote{http://sites.nd.edu/nethealth/}\cite{PurtaNetHealth}, which was collected from 698 students in total during eight semesters. This comprehensive study includes data on communication patterns from mobile phones, sleep, and physical activity routines, students' family backgrounds, living conditions, personality, etc., from surveys. This study uses some parts of the basic survey and all the wearable device data. Therefore, to extract the required data, preprocessing is applied to construct a sub-dataset for the analysis.

In the original NetHealth dataset, several sub-components, such as communication, wearable, survey, courses and grades, and calendar are included. Wearable measurements are performed for activity and sleep tracking using a Fitbit device. We used wearable and survey datasets for this study to construct a sub-dataset to apply deep learning and ensemble learning. However, this raw data form is incompatible with training deep learning algorithms. The preprocessing step will be further discussed in the paper. 

We applied several CNN architectures which perform well in the human activity recognition domain~\cite{zhang2022deep}. As our measurements in the dataset are collected from wearables on a daily basis, we decided to focus on the architectures utilized for daily activity recognition tasks. Details of these architectures and their differences are further discussed in the paper. We also applied Random Forest (RF) as an ensemble learning method to compare the performance with deep learning architectures. RF is also used to extract important features for sleep quality prediction. Even though deep learning algorithms, especially CNNs, perform better compared to conventional machine learning techniques, our results show that in this context, RF performs better. We believe that it is highly related to dataset characteristics since in \cite{booth2022toward}, authors also applied different deep learning techniques such as Multilayer Perceptron (MLP), Long-Short Term Memory Network (LSTM), Gated Recurrent Unit (GRU), Elastic Net (EN), and compared them with conventional machine learning technique; RF. They found RF to be performing better than other methods. 

The rest of the paper is organized as follows: Section 2 explains state of the art on sleep recognition studies from the point of the wearable domain. Section 3 explains dataset details and the preprocessing steps for further analyses. In Section 4, we present model results with different proposed architectures. Finally, Section 5 discusses our findings with other future study ideas.

\section{Related Works}
 Several related studies exist about the sleep quality prediction of people that use wearable devices and questionnaires as measurement metrics. We can emphasize the studies \cite{kilicc2022prediction, phan2020applying, sathyanarayana2016sleep, sano2015recognizing,  arora2020analysis, 9202634, PurtaNetHealth} as the most related ones to our study. Although the personal survey is not included in all the mentioned works, wearable devices are the key for all of them.

In \cite{arora2020analysis}, authors use three different sleep quality metrics: Daily Sleep Quality, Weekly Sleep Quality, and Sleep Consistency. CNN and MLP methods are applied. CNN is found to be the outperforming method compared to MLP on all these three quality metrics. There were three classes for daily and weekly sleep quality metrics indicating good to poor quality; and four classes for sleep consistency metric which is obtained by the weighted sum of the standard deviations of sleep, rest, bedtime, out-of-bedtime, and sleep inconsistency between weekdays and weekends. CNN obtains the best results on weekly sleep quality classification with $97.30\%$ accuracy.

In \cite{9202634}, the main aim was correctly predicting the future sleep duration, which will help identify future sleep quality. As a model, they utilized a generalized linear mixed model (GLMM) to consider individual variability on sleep duration, which provides better estimates, hence better sleep quality predictions, in addition to a generalized linear model (GLM) for considering fixed effects.

In \cite{kilicc2022prediction}, sleep and activity metrics are used to estimate university students' well-being. The same data set is used in this study, NetHealth data. Daily sleep and activity data is used to train a classification algorithm. The study aims to predict the ``Subjective Well-being" score of people from average daily steps, average heart rate, heartbeat standard deviation, average sleep duration, and sleep duration deviation gathered from wearable devices. They used Naive Bayes, K-Nearest Neighbors (KNN), Support Vector Machine (SVM), and Ensemble classifiers as methods to train.


In \cite{phan2020applying}, data from 39 students using Fitbit devices is collected during 106 days. They focused on predicting two classes: good quality and poor quality of sleep based on seven features, such as calories, steps, distance, sedentary, lightly active, fairly active, very active. They applied Random Forest as a conventional machine learning method and LSTM, GRU, and CNN as deep learning methods. Their obtained results indicate that CNN could improve sleep quality prediction from activities. But the best performing model is reported as GRU.

In \cite{PurtaNetHealth}, authors discuss the compliance of Fitbit usage among college students and the reasons for non-compliance. They observed patterns for general health, sleep, and exercise. Mainly, a contrastive pattern between weekdays and weekends, regular school days, and deadline-intensive patterns are observed among the ones with a compliance rate higher than $60\%$.

 Study \cite{sano2015recognizing} has a wide range of target values while using personality traits, wearable device data, and mobile phone data. They collected multi-modal data from 66 people for a month. This multi-modal data includes perceived stress, sleep, personality, physiological, behavioral, and social interaction data to train a model to divine academic performance, sleep, stress, and mental health scores. They used SVM-L and SVM-RBF models to train and compared the GPA (Academic Preformance), PSQI (Pittsburg Sleep Quality Index), PSS (Perceived Stress Scale), and MCS (Mental Health Composite Score) results for different training datasets. As sleep was not a target in the scope of this study, they examined only its relation with other terms. It is found that poor sleep quality and high-stress levels are highly correlated. In addition, the poor sleep quality group has been found to have more screen time around 3-6 am and less around 9 am-12 pm. 
 
In \cite{sathyanarayana2016sleep}, the aim is to classify sleep quality with a deep learning model trained with wearable device data. They used the physical activity data collected during awake time. The target values of the prediction model are poor or good sleep quality. The wearable measurements are done using an Actigraphy device which is accepted as a gold-standard device for clinical studies of sleep and physical activity patterns. Data is collected from 92 adolescents over one whole week. Their methods are traditional logistic regression as a conventional ML method and more advanced deep learning methods: multilayer perceptron (MLP), convolutional neural network (CNN), simple Elman-type recurrent neural network (RNN), long short-term memory (LSTM-RNN), and a time-batched version of LSTM-RNN (TB-LSTM). At the end of the study, they revealed that the CNN has the highest accuracy with 0.9449 while the traditional logistic regression has a score of 0.6463 accuracies. 

Our study differs from the literature with its extensive feature sets and its nature to predict sleep efficiency, contrary to already applied classification techniques. We contribute to the literature by applying CNN architectures which already performed well in the human activity recognition domain. We also compare the performance with an ensemble model, i.e., RF. It is found that RF performs better compared to CNN architectures in the scope of our experiments.


\section{Methodology}
\subsection{Dataset and Features}
The data used in this study is gathered from the Nethealth dataset, which is collected at Notre Dame University. This data covers eight semesters between Fall 2015 and Spring 2019. There are data on approximately 400 students from 2015 to 2017 and 300 from 2015 to 2019. Data collection comprises social networks, physical activity, sleep data, basic survey data, communication data, courses and grades data, and academic calendar data. Basic survey data include family background \& demographic variables, student background, self-reported course, grade, major information, activities, clubs, musical tastes, personality - Big five, and social-psychological scales tests. Besides these, physical and psychological health-related questions, PSQI sleep quality test, political attributes and views, and computer, phone, and Fitbit device usage of students are included.

NetHealth dataset is a comprehensive, multi-year longitudinal dataset. So, some survey questions were only asked in some wave. The missing information restricts this study from using every wave in deep learning training. In this study, we focus on Fitbit's sleep quality scores as our target value. 

In the scope of this study, we chose all wearable data modalities, i.e., activity and sleep. However, wearable sleep data is chosen as the target value. Thus, it is not in the feature space. In addition, we added questionnaire responses about bad habits, personality (BigFive), exercise, health, mental health, personal information, origin, and sex. The used sub-datasets and corresponding features are listed in study \cite{saylam2022academic} in Table 1. Due to space limit we were not able to add the adjusted version in this paper. However, the parameter space remains the same but removed the courses and grades features. There, the underscored numbers correspond to the semester's indications. The feature space size before pre-processing is $89$ without participantID, date, and the target value, which is sleep efficiency.

\begin{figure*}
  \includegraphics[width=\textwidth]{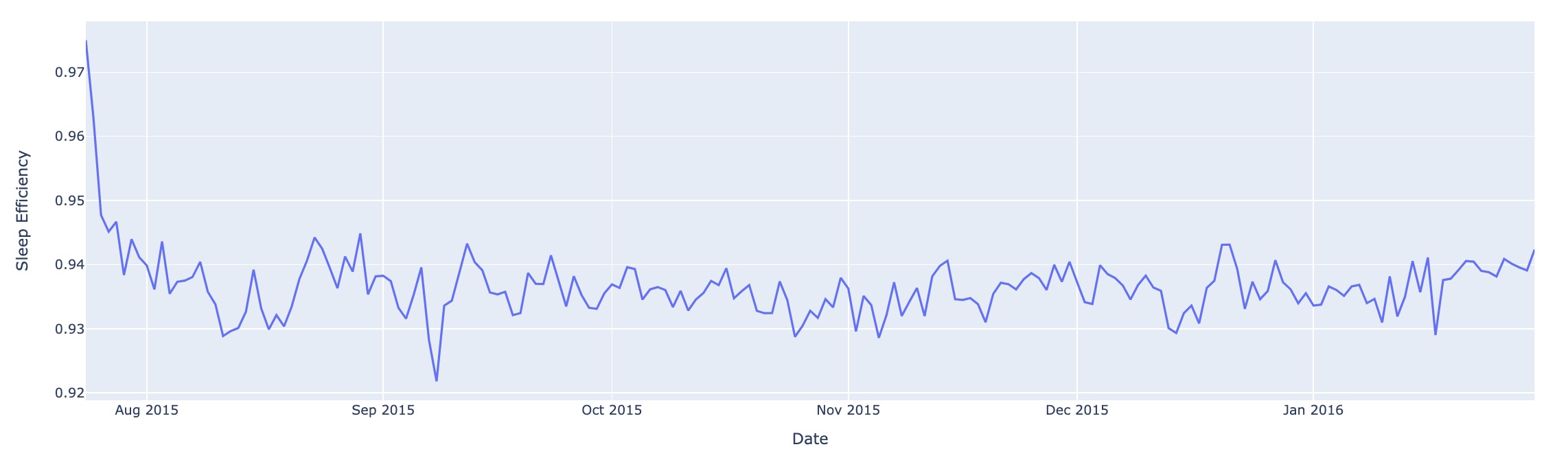}
  \caption{Sleep Efficiency for first and second semesters}
  \Description{Sleep Quality over 2 semesters \textcolor{red}{Refer to this figure in the text. What is it? explain. Why are there drops sometimes?}}
  \label{fig:SleepAllWaves}
\end{figure*}

\subsection{Preprocessing}
As a target value in the scope of this study, we use the Fitbit sleep efficiency score that is collected from the wearable - Fitbit. The device uses its sensors to detect if someone is awake or asleep. Besides this information, total time in bed, awake time in bed, and sleeping time are also detected. Thanks to these values, the sleep efficiency score can be calculated, i.e., $minsasleep/(minsasleep + minsawake)$. The score is the ratio of asleep time in bed to the total time in bed and ranges between $0$ and $1$. 

Before model evaluation, we concatenated all separated responses according to timestamps and participants. In addition, we converted the survey questions into numeric models. Besides, there were missing values in some survey responses for different semesters. In this study, we are concentrating on the first and second semester's data due to not being requested from the participants since the decided features have responses. Furthermore, students who still need to answer the questionnaires are subtracted. In the end, we have $200615$ rows and $92$ columns. In Figure \ref{fig:SleepAllWaves}, the change of our target variable, i.e., sleep efficiency, over the first and second semesters can be observed. It consists of all participants' overall patterns in two semesters. Its range is between $0.93$ and $0.95$. There were two peaks; one at the beginning of the study in August 2015 and the other around September 2015. However, we can not comment on the possible reasons since we do not know about the events that occur these days.

\subsection{Model Details and Performance Metrics} 
In this study, we apply deep learning techniques as they perform better in literature \cite{zhang2022deep}. More specifically, we are concentrating on Convolutional Neural Networks (CNN) architectures that perform well in the human activity recognition domain \cite{zeng2014convolutional, triwiyanto2020improved, xu2021human, ha2016convolutional, jordao2018novel}. As our measurements in the dataset are collected daily from wearables, we decided to focus on the architectures utilized for daily activity recognition tasks. The selected architectures are listed in Table \ref{table:CnnTable}. C, P, and FC stand for convolution layer, max-pooling layer, and fully connected layer, respectively. We applied these selected architectures to both scenarios. The models differ from each other primarily on their complexities, and also, we chose different kernel sizes already performed well, which are $3, 5, 12, 20$. We used ReLU activation for convolution and linear activation for fully connected layers. Filter and pool size remained the same for all architectures, 32 and 2, respectively.

Since our problem is prediction, mean squared error (MSE) and mean absolute error (MAE) can be utilized as performance metrics. MAE is calculated by considering the absolute average distance. MSE is calculated by averaging the squared difference between the estimated and actual values. In our context, we used MAE due to its interpretability.


\begin{table}[!htb]
  \begin{center}
    \caption{Details of Model Parameters}
     \label{table:CnnTable}
    \begin{tabular}{l|rrrrrrr} 
    \hline
      \textbf{Architecture} & \textbf{Layer} & \textbf{Activation} & \textbf{Kernel Size} & \textbf{Pool Size} & \textbf{Filters} & \textbf{Input Size} & \textbf{Output Size}\\
      \hline
      C-P-FC-FC  & 1D Conv. & ReLU & 20 & - & 32 & (93, 1) & (74, 32)\\

       & Pooling & - & - & 2 & - & - & (37, 32) \\
        & Dense & Linear & - & - & - & (93, 1) & (37, 16)\\
       & Dense & Linear & - & - & - & - & (37, 16)\\
       & Flatten & - & - & - & - & - & 592\\
     \hline
 C-P-C-P-FC   & 1D Conv. & ReLU & 3 & - & 32 & (93, 1) & (91, 32)\\
& Pooling & - & - & 2 & - & - & (45, 32) \\
& 1D Conv. & ReLU & 3 & - & 32 & - & (43, 32)\\
 & Pooling & - & - & 2 & - & - & (21, 32) \\
 & Dense & Linear & - & - & - & (50, 1) & (21, 16)\\
  & Flatten & - & - & - & - & - & 336\\
     \hline
     C-P-C-P-FC   & 1D Conv. & ReLU & 5 & - & 32 & (93, 1) & (89, 32)\\
  & Pooling & - & - & 2 & - & - & (44, 32) \\
& 1D Conv. & ReLU & 5 & - & 32 & - & (40, 32)\\
 & Pooling & - & - & 2 & - & - & (20, 32) \\
 & Dense & Linear & - & - & - & (50, 1) & (20, 16)\\
  & Flatten & - & - & - & - & - & 320\\
      \hline
C-C-P-C-C-P-FC  & 1D Conv. & ReLU & 5 & - & 32 & (93, 1) & (89, 32)\\
& 1D Conv. & ReLU & 5 & - & 32 & - & (85, 32)\\
 & Pooling & - & - & 2 & - & - & (42, 32) \\
 & 1D Conv. & ReLU & 5 & - & 32 & - & (38, 32)\\
 & 1D Conv. & ReLU & 5 & - & 32 & - & (34, 32)\\
 & Pooling & - & - & 2 & - & - & (17, 32) \\
& Dense & Linear & - & - & - & (93, 1) & (17, 16)\\
  & Flatten & - & - & - & - & - & 272\\
  
       \hline
C-P-C-P-FC-FC  & 1D Conv. & ReLU & 5 & - & 32 & (93, 1) & (89, 32)\\
 & Pooling & - & - & 2 & - & - & (44, 32) \\
 & 1D Conv. & ReLU & 5 & - & 32 & - & (40, 32)\\
 & Pooling & - & - & 2 & - & - & (20, 32) \\
 & Dense & Linear & - & - & - & (93, 1) & (20, 16)\\
 & Dense & Linear & - & - & - & - & (20, 16)\\
 & Flatten & - & - & - & - & - & 320\\
      \hline
     C-P-C-P-C-P   & 1D Conv. & ReLU & 12 & - & 32 & (93, 1) & (82, 32)\\
   & Pooling & - & - & 2 & - & - & (41, 32) \\
& 1D Conv. & ReLU & 12 & - & 32 & - & (30, 32)\\
 & Pooling & - & - & 2 & - & - & (15, 32) \\
 & 1D Conv. & ReLU & 12 & - & 32 & - & (4, 32)\\
 & Pooling & - & - & 2 & - & - & (2, 32) \\
 & Flatten & - & - & - & - & - & 64\\
 \hline

    \end{tabular}
  \end{center}
\end{table}

\section{Evaluation}
\subsection{Comparison of CNN models and Random Forest }
In the dataset, we have continuous measurements for sleep efficiency. Thus, we modelled them as a prediction problem. Again, we employed the same architectures, and obtained results are presented in Table \ref{table:sleepPrediction}. Architecture 2 is the best performer based on MAE values. The change of MAE for the best one (C-P-C-P-FC) is given in Figure \ref{fig:MSE3and4}. In addition, we implemented RF with $10$ estimators; its performance was found to be better than the CNN models. We think that this may be related to the dataset characteristics since in \cite{booth2022toward}, RF outperformed different deep learning models on the same dataset but on a different task.

\begin{minipage}{\textwidth}
  \begin{minipage}[b]{0.49\textwidth}
   \centering
    \includegraphics[width=.85\textwidth]{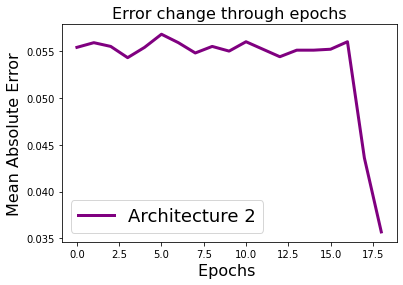}
    \captionof{figure}{Evaluation of the best architecture}
    \label{fig:MSE3and4}
    
  \end{minipage}
  \hfill
  \begin{minipage}[b]{0.49\textwidth}

    \centering
  
    \begin{tabular}{cc}\hline
      \textbf{Architecture} & \textbf{Mean Absolute Error} \\ \hline
        C-P-FC-FC         &     0.0876              \\ 
        C-P-C-P-FC        &    \textbf{0.0357}            \\
        C-P-C-P-FC        &  0.0429  \\
        C-C-P-C-C-P-FC &  0.0844 \\
        C-P-C-P-FC-FC     & 0.0679 \\     
        C-P-C-P-C-P & 0.9361\\ \hline
        Random Forest & \textbf{0.0282}  \\
        \hline 
      \end{tabular}
      \captionof{table}{Sleep Quality Prediction Results}
      \label{table:sleepPrediction}
       
    \end{minipage}
  \end{minipage}

\subsection{Important Factors}
\begin{figure}[!t]
  \includegraphics[width=0.7\textwidth]{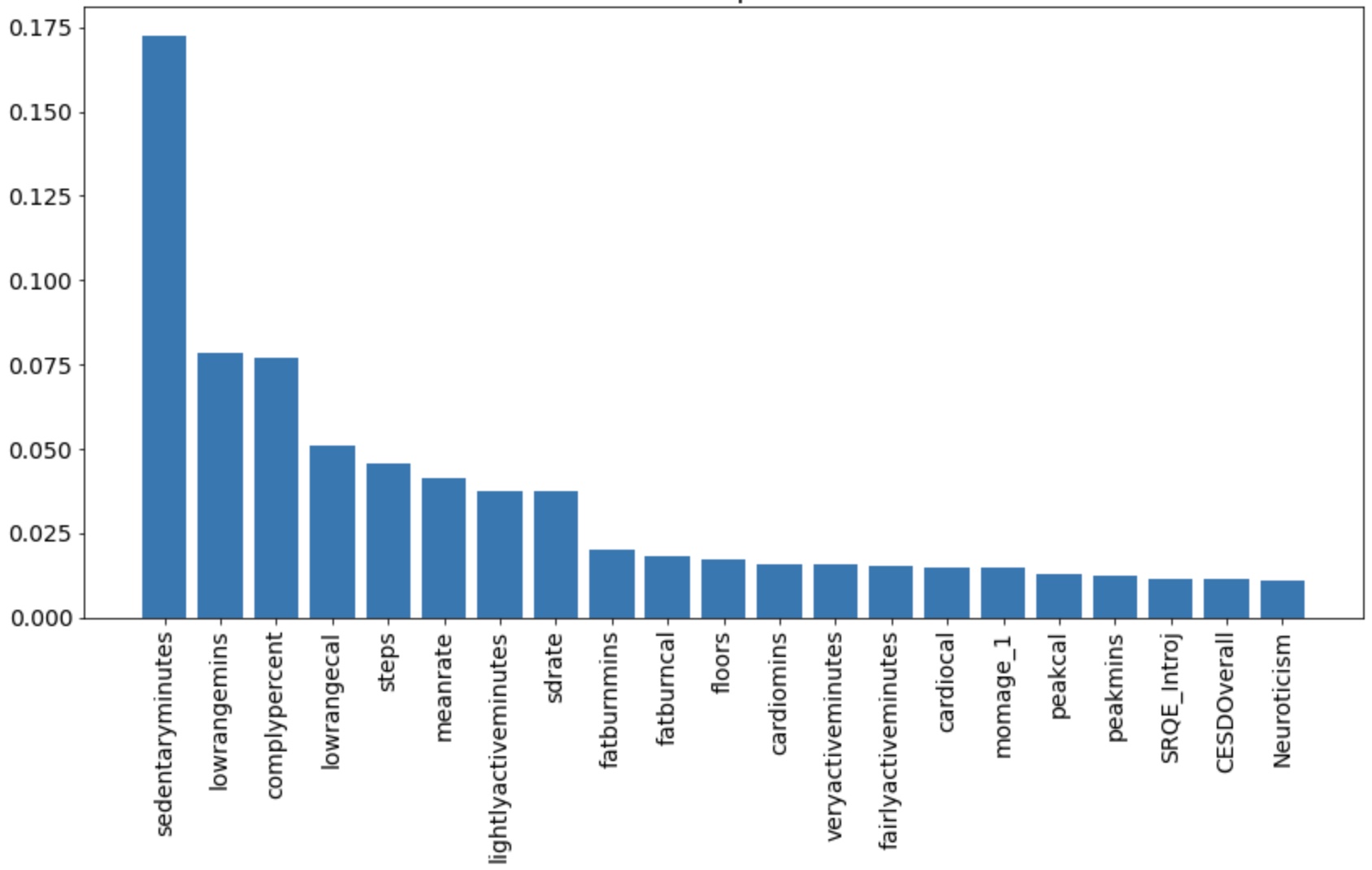}
  \caption{Most Important Factors for Sleep Quality Prediction}
  \Description{Important Factors}
  \label{fig:ImportantFactors}
\end{figure}

To understand the most affecting factors to sleep efficiency, we employed RF feature importance. $20$ top affecting ones are presented in Figure \ref{fig:ImportantFactors}. They are mostly related to Fitbit activity measurements. We got SRQE\_introj (introjective self-regulation for exercise), CESDOverall (CES depression score), Neuroticism (personality trait) from survey responses. Also, we found mom age affects sleep efficiency somehow. 

We re-ran the indicated architectures using only the selected most important factors. However, we got higher MAE values compared to utilizing all factors. We can conclude that in this context even though the final performance contribution is low for some features, when we exclude them, the overall prediction performance decreases.

\section{Discussion and Conclusion}
In this study, we predict sleep efficiency values collected from Fitbit devices. We applied the most promising CNN architectures in the human activity domain. We also implemented conventional RF to be able to compare deep learning results. We also extracted the most affecting factors using the feature importance technique. 

Even though CNN predictions are promising and the best architecture is found as C-P-C-P-FC, RF performs the best in our context but the difference is small in terms of MAE. This is probably due to the nature of the dataset since in \cite{booth2022toward}, authors applied different deep learning techniques, and it was found that RF performs the best in all cases. 

We also experimented with the reduced size of features in the same scenario. However, it is found that deep learning without feature extraction performs better compared to manually extracted features. Nevertheless, this feature ranking provides us with a meaning for the interpretation. Among the best $20$ factors, sleep efficiency is affected by the measurements from wearables related to the activity. There are also some survey-related factors, such as exercise, depression, and personality-related responses. These are expected since, in literature, it is already shown that there is a relation between sleep, stress, physical activity, and mood \cite{marcusson2022relationships, gedam2021review, steptoe2008positive, pilcher1996effects}. We contribute by working all these modalities in one study in a multi-modal manner by applying deep learning methods over wearable and survey data collected in an unrestricted environment.

We want to give ideas for future perspectives. In this study, we worked on a prediction problem, but the target value range was narrow (between $0.93$ and $0.95$). There were no high fluctuations. This may be the reason for obtaining very accurate predictions and low errors. In addition, we focused on the application of algorithms by not considering personal differences. This aspect may be examined in future studies. Furthermore, we did not consider the time-based differences for a participant. One may examine personal changes over time periods, especially during exam dates and holidays.







\section{Acknowledgments}
Tübitak Bideb 2211-A academic reward is gratefully acknowledged. This research has been supported by the Boğaziçi University Research Fund, project number: 19301P. 
\bibliography{sample-base}


\begin{thebibliography}{19}


\ifx \showCODEN    \undefined \def \showCODEN     #1{\unskip}     \fi
\ifx \showDOI      \undefined \def \showDOI       #1{#1}\fi
\ifx \showISBNx    \undefined \def \showISBNx     #1{\unskip}     \fi
\ifx \showISBNxiii \undefined \def \showISBNxiii  #1{\unskip}     \fi
\ifx \showISSN     \undefined \def \showISSN      #1{\unskip}     \fi
\ifx \showLCCN     \undefined \def \showLCCN      #1{\unskip}     \fi
\ifx \shownote     \undefined \def \shownote      #1{#1}          \fi
\ifx \showarticletitle \undefined \def \showarticletitle #1{#1}   \fi
\ifx \showURL      \undefined \def \showURL       {\relax}        \fi
\providecommand\bibfield[2]{#2}
\providecommand\bibinfo[2]{#2}
\providecommand\natexlab[1]{#1}
\providecommand\showeprint[2][]{arXiv:#2}

\bibitem[Arora et~al\mbox{.}(2020)]%
        {arora2020analysis}
\bibfield{author}{\bibinfo{person}{Anshika Arora}, \bibinfo{person}{Pinaki
  Chakraborty}, {and} \bibinfo{person}{MPS Bhatia}.}
  \bibinfo{year}{2020}\natexlab{}.
\newblock \showarticletitle{Analysis of data from wearable sensors for sleep
  quality estimation and prediction using deep learning}.
\newblock \bibinfo{journal}{\emph{Arabian Journal for Science and Engineering}}
  \bibinfo{volume}{45}, \bibinfo{number}{12} (\bibinfo{year}{2020}),
  \bibinfo{pages}{10793--10812}.
\newblock


\bibitem[Booth et~al\mbox{.}(2022)]%
        {booth2022toward}
\bibfield{author}{\bibinfo{person}{Brandon~M Booth}, \bibinfo{person}{Hana
  Vrzakova}, \bibinfo{person}{Stephen~M Mattingly}, \bibinfo{person}{Gonzalo~J
  Martinez}, \bibinfo{person}{Louis Faust}, {and} \bibinfo{person}{Sidney~K
  D’Mello}.} \bibinfo{year}{2022}\natexlab{}.
\newblock \showarticletitle{Toward Robust Stress Prediction in the Age of
  Wearables: Modeling Perceived Stress in a Longitudinal Study with Information
  Workers}.
\newblock \bibinfo{journal}{\emph{IEEE Transactions on Affective Computing}}
  \bibinfo{volume}{13}, \bibinfo{number}{4} (\bibinfo{year}{2022}),
  \bibinfo{pages}{2201--2217}.
\newblock


\bibitem[Chen et~al\mbox{.}(2020)]%
        {9202634}
\bibfield{author}{\bibinfo{person}{Chih-You Chen}, \bibinfo{person}{Sudip
  Vhaduri}, {and} \bibinfo{person}{Christian Poellabauer}.}
  \bibinfo{year}{2020}\natexlab{}.
\newblock \showarticletitle{Estimating Sleep Duration from Temporal Factors,
  Daily Activities, and Smartphone Use}. In \bibinfo{booktitle}{\emph{2020 IEEE
  44th Annual Computers, Software, and Applications Conference (COMPSAC)}}.
  \bibinfo{pages}{545--554}.
\newblock
\urldef\tempurl%
\url{https://doi.org/10.1109/COMPSAC48688.2020.0-196}
\showDOI{\tempurl}


\bibitem[Gedam and Paul(2021)]%
        {gedam2021review}
\bibfield{author}{\bibinfo{person}{Shruti Gedam} {and}
  \bibinfo{person}{Sanchita Paul}.} \bibinfo{year}{2021}\natexlab{}.
\newblock \showarticletitle{A review on mental stress detection using wearable
  sensors and machine learning techniques}.
\newblock \bibinfo{journal}{\emph{IEEE Access}}  \bibinfo{volume}{9}
  (\bibinfo{year}{2021}), \bibinfo{pages}{84045--84066}.
\newblock


\bibitem[Ha and Choi(2016)]%
        {ha2016convolutional}
\bibfield{author}{\bibinfo{person}{Sojeong Ha} {and} \bibinfo{person}{Seungjin
  Choi}.} \bibinfo{year}{2016}\natexlab{}.
\newblock \showarticletitle{Convolutional neural networks for human activity
  recognition using multiple accelerometer and gyroscope sensors}. In
  \bibinfo{booktitle}{\emph{2016 international joint conference on neural
  networks (IJCNN)}}. IEEE, \bibinfo{pages}{381--388}.
\newblock


\bibitem[Jordao et~al\mbox{.}(2018)]%
        {jordao2018novel}
\bibfield{author}{\bibinfo{person}{Artur Jordao}, \bibinfo{person}{Leonardo
  Ant{\^o}nio~Borges Torres}, {and} \bibinfo{person}{William~Robson Schwartz}.}
  \bibinfo{year}{2018}\natexlab{}.
\newblock \showarticletitle{Novel approaches to human activity recognition
  based on accelerometer data}.
\newblock \bibinfo{journal}{\emph{Signal, Image and Video Processing}}
  \bibinfo{volume}{12} (\bibinfo{year}{2018}), \bibinfo{pages}{1387--1394}.
\newblock


\bibitem[K{\i}l{\i}{\c{c}} et~al\mbox{.}(2022)]%
        {kilicc2022prediction}
\bibfield{author}{\bibinfo{person}{Akif~Can K{\i}l{\i}{\c{c}}},
  \bibinfo{person}{Ahmet Karaku{\c{s}}}, {and} \bibinfo{person}{Emre
  Alptekin}.} \bibinfo{year}{2022}\natexlab{}.
\newblock \showarticletitle{Prediction of University Students’ Subjective
  Well-Being with Sleep and Physical Activity Data using Classification
  Algorithms}.
\newblock \bibinfo{journal}{\emph{Procedia Computer Science}}
  \bibinfo{volume}{207} (\bibinfo{year}{2022}), \bibinfo{pages}{2648--2657}.
\newblock


\bibitem[Marcusson-Clavertz et~al\mbox{.}(2022)]%
        {marcusson2022relationships}
\bibfield{author}{\bibinfo{person}{David Marcusson-Clavertz},
  \bibinfo{person}{Martin~J Sliwinski}, \bibinfo{person}{Orfeu~M Buxton},
  \bibinfo{person}{Jinhyuk Kim}, \bibinfo{person}{David~M Almeida}, {and}
  \bibinfo{person}{Joshua~M Smyth}.} \bibinfo{year}{2022}\natexlab{}.
\newblock \showarticletitle{Relationships between daily stress responses in
  everyday life and nightly sleep}.
\newblock \bibinfo{journal}{\emph{Journal of behavioral medicine}}
  \bibinfo{volume}{45}, \bibinfo{number}{4} (\bibinfo{year}{2022}),
  \bibinfo{pages}{518--532}.
\newblock


\bibitem[Phan et~al\mbox{.}(2020)]%
        {phan2020applying}
\bibfield{author}{\bibinfo{person}{Dinh-Van Phan}, \bibinfo{person}{Chien-Lung
  Chan}, {and} \bibinfo{person}{Duc-Khanh Nguyen}.}
  \bibinfo{year}{2020}\natexlab{}.
\newblock \showarticletitle{Applying Deep Learning for Prediction Sleep Quality
  from Wearable Data}. In \bibinfo{booktitle}{\emph{Proceedings of the 4th
  International Conference on Medical and Health Informatics}}.
  \bibinfo{pages}{51--55}.
\newblock


\bibitem[Pilcher and Huffcutt(1996)]%
        {pilcher1996effects}
\bibfield{author}{\bibinfo{person}{June~J Pilcher} {and}
  \bibinfo{person}{Allen~I Huffcutt}.} \bibinfo{year}{1996}\natexlab{}.
\newblock \showarticletitle{Effects of sleep deprivation on performance: a
  meta-analysis}.
\newblock \bibinfo{journal}{\emph{Sleep}} \bibinfo{volume}{19},
  \bibinfo{number}{4} (\bibinfo{year}{1996}), \bibinfo{pages}{318--326}.
\newblock


\bibitem[Purta et~al\mbox{.}(2016)]%
        {PurtaNetHealth}
\bibfield{author}{\bibinfo{person}{Rachael Purta}, \bibinfo{person}{Stephen
  Mattingly}, \bibinfo{person}{Lixing Song}, \bibinfo{person}{Omar Lizardo},
  \bibinfo{person}{David Hachen}, \bibinfo{person}{Christian Poellabauer},
  {and} \bibinfo{person}{Aaron Striegel}.} \bibinfo{year}{2016}\natexlab{}.
\newblock \showarticletitle{Experiences Measuring Sleep and Physical Activity
  Patterns across a Large College Cohort with Fitbits}. In
  \bibinfo{booktitle}{\emph{Proceedings of the 2016 ACM International Symposium
  on Wearable Computers}} (Heidelberg, Germany) \emph{(\bibinfo{series}{ISWC
  '16})}. \bibinfo{publisher}{Association for Computing Machinery},
  \bibinfo{address}{New York, NY, USA}, \bibinfo{pages}{28–35}.
\newblock
\showISBNx{9781450344609}
\urldef\tempurl%
\url{https://doi.org/10.1145/2971763.2971767}
\showDOI{\tempurl}


\bibitem[Sano et~al\mbox{.}(2015)]%
        {sano2015recognizing}
\bibfield{author}{\bibinfo{person}{Akane Sano}, \bibinfo{person}{Andrew~J
  Phillips}, \bibinfo{person}{Z~Yu Amy}, \bibinfo{person}{Andrew~W McHill},
  \bibinfo{person}{Sara Taylor}, \bibinfo{person}{Natasha Jaques},
  \bibinfo{person}{Charles~A Czeisler}, \bibinfo{person}{Elizabeth~B Klerman},
  {and} \bibinfo{person}{Rosalind~W Picard}.} \bibinfo{year}{2015}\natexlab{}.
\newblock \showarticletitle{Recognizing academic performance, sleep quality,
  stress level, and mental health using personality traits, wearable sensors
  and mobile phones}. In \bibinfo{booktitle}{\emph{2015 IEEE 12th International
  Conference on Wearable and Implantable Body Sensor Networks (BSN)}}. IEEE,
  \bibinfo{pages}{1--6}.
\newblock


\bibitem[Sathyanarayana et~al\mbox{.}(2016)]%
        {sathyanarayana2016sleep}
\bibfield{author}{\bibinfo{person}{Aarti Sathyanarayana},
  \bibinfo{person}{Shafiq Joty}, \bibinfo{person}{Luis Fernandez-Luque},
  \bibinfo{person}{Ferda Ofli}, \bibinfo{person}{Jaideep Srivastava},
  \bibinfo{person}{Ahmed Elmagarmid}, \bibinfo{person}{Teresa Arora},
  \bibinfo{person}{Shahrad Taheri}, {et~al\mbox{.}}}
  \bibinfo{year}{2016}\natexlab{}.
\newblock \showarticletitle{Sleep quality prediction from wearable data using
  deep learning}.
\newblock \bibinfo{journal}{\emph{JMIR mHealth and uHealth}}
  \bibinfo{volume}{4}, \bibinfo{number}{4} (\bibinfo{year}{2016}),
  \bibinfo{pages}{e6562}.
\newblock


\bibitem[Saylam et~al\mbox{.}(2022)]%
        {saylam2022academic}
\bibfield{author}{\bibinfo{person}{Berrenur Saylam},
  \bibinfo{person}{Ekrem~Yusuf Ekmekci}, \bibinfo{person}{Eren
  Altuno{\u{g}}lu}, {and} \bibinfo{person}{Ozlem~Durmaz Incel}.}
  \bibinfo{year}{2022}\natexlab{}.
\newblock \showarticletitle{Academic Performance Relation with Behavioral
  Trends and Personal Characteristics: Wearable Device Perspective}.
\newblock  (\bibinfo{year}{2022}).
\newblock


\bibitem[Steptoe et~al\mbox{.}(2008)]%
        {steptoe2008positive}
\bibfield{author}{\bibinfo{person}{Andrew Steptoe}, \bibinfo{person}{Katie
  O'Donnell}, \bibinfo{person}{Michael Marmot}, {and} \bibinfo{person}{Jane
  Wardle}.} \bibinfo{year}{2008}\natexlab{}.
\newblock \showarticletitle{Positive affect, psychological well-being, and good
  sleep}.
\newblock \bibinfo{journal}{\emph{Journal of psychosomatic research}}
  \bibinfo{volume}{64}, \bibinfo{number}{4} (\bibinfo{year}{2008}),
  \bibinfo{pages}{409--415}.
\newblock


\bibitem[Triwiyanto et~al\mbox{.}(2020)]%
        {triwiyanto2020improved}
\bibfield{author}{\bibinfo{person}{Triwiyanto Triwiyanto},
  \bibinfo{person}{I~Putu~Alit Pawana}, {and} \bibinfo{person}{Mauridhi~Hery
  Purnomo}.} \bibinfo{year}{2020}\natexlab{}.
\newblock \showarticletitle{An improved performance of deep learning based on
  convolution neural network to classify the hand motion by evaluating hyper
  parameter}.
\newblock \bibinfo{journal}{\emph{IEEE Transactions on Neural Systems and
  Rehabilitation Engineering}} \bibinfo{volume}{28}, \bibinfo{number}{7}
  (\bibinfo{year}{2020}), \bibinfo{pages}{1678--1688}.
\newblock


\bibitem[Xu and Qiu(2021)]%
        {xu2021human}
\bibfield{author}{\bibinfo{person}{Yang Xu} {and} \bibinfo{person}{Ting~Ting
  Qiu}.} \bibinfo{year}{2021}\natexlab{}.
\newblock \showarticletitle{Human activity recognition and embedded application
  based on convolutional neural network}.
\newblock \bibinfo{journal}{\emph{Journal of Artificial Intelligence and
  Technology}} \bibinfo{volume}{1}, \bibinfo{number}{1} (\bibinfo{year}{2021}),
  \bibinfo{pages}{51--60}.
\newblock


\bibitem[Zeng et~al\mbox{.}(2014)]%
        {zeng2014convolutional}
\bibfield{author}{\bibinfo{person}{Ming Zeng}, \bibinfo{person}{Le~T Nguyen},
  \bibinfo{person}{Bo Yu}, \bibinfo{person}{Ole~J Mengshoel},
  \bibinfo{person}{Jiang Zhu}, \bibinfo{person}{Pang Wu}, {and}
  \bibinfo{person}{Joy Zhang}.} \bibinfo{year}{2014}\natexlab{}.
\newblock \showarticletitle{Convolutional neural networks for human activity
  recognition using mobile sensors}. In \bibinfo{booktitle}{\emph{6th
  international conference on mobile computing, applications and services}}.
  IEEE, \bibinfo{pages}{197--205}.
\newblock


\bibitem[Zhang et~al\mbox{.}(2022)]%
        {zhang2022deep}
\bibfield{author}{\bibinfo{person}{Shibo Zhang}, \bibinfo{person}{Yaxuan Li},
  \bibinfo{person}{Shen Zhang}, \bibinfo{person}{Farzad Shahabi},
  \bibinfo{person}{Stephen Xia}, \bibinfo{person}{Yu Deng}, {and}
  \bibinfo{person}{Nabil Alshurafa}.} \bibinfo{year}{2022}\natexlab{}.
\newblock \showarticletitle{Deep learning in human activity recognition with
  wearable sensors: A review on advances}.
\newblock \bibinfo{journal}{\emph{Sensors}} \bibinfo{volume}{22},
  \bibinfo{number}{4} (\bibinfo{year}{2022}), \bibinfo{pages}{1476}.
\newblock


\end{thebibliography}

\end{document}